\documentclass[iop,apj]{emulateapj}
\slugcomment{To appear in the Astrophysical Journal}

\begin{document}

\title{Protoplanetary Disk Masses in IC348: A Rapid Decline in the Population of Small Dust Grains After 1 Myr}
\author{Nicholas Lee, Jonathan P. Williams, Lucas A. Cieza}
\affil{Institute for Astronomy, 2680 Woodlawn Dr., Honolulu, HI, 96822, USA}
\shorttitle{Protoplanetary disks in IC348}
\shortauthors{Lee et al.}

\begin{abstract}
We present a 1.3\,mm continuum survey of protoplanetary disks in
the 2--3\,Myr old cluster, IC348, with the Submillimeter Array.
We observed 85 young stellar objects and
detected 10 with 1.3\,mm fluxes greater than 2\,mJy.
The brightest source is a young embedded protostar
driving a molecular outflow. The other 9 detections are dusty
disks around optically visible stars. Our millimeter flux
measurements translate into total disk masses ranging from 2 to 6
Jupiter masses. Each detected disk has strong mid-infrared
emission in excess of the stellar photosphere and has H$\alpha$
equivalent widths larger than the average in the cluster and
indicative of ongoing gas accretion.
The disk mass distribution, however, is shifted by about a factor of
20 to lower masses, compared to that in the $\sim 1$\,Myr old 
Taurus and Ophiuchus regions. These observations reveal the rapid
decline in the number of small dust grains in disks with time,
and probably their concomitant growth beyond millimeter sizes.
Moreover, if IC348 is to form planets in the same proportion as
detected in the field, these faint millimeter detections may represent
the best candidates in the cluster to study the progression from
planetesimals to planets.
\end{abstract}


\section{Introduction}
Dusty disks persist around most low mass stars for a Myr or more,
much longer than the lifetime of the natal molecular core.
At the stage where the core is dispersed, the now optically visible
central star has nearly achieved its final mass and the disk is
no longer protostellar, but it may still be protoplanetary.
In the literature of young stellar objects (YSOs), this corresponds to
physical stage II \citep{2006ApJS..167..256R}
and observational Class II \citep{1987IAUS..115....1L}.

The presence of protoplanetary disks is most readily inferred
via optically thick infrared emission from the warm dust
close to the star. The capacity of a disk to actually form planets,
however, requires knowledge of its surface density (or at least mass)
which necessitates observations at millimeter wavelengths where the
thermal dust emission is optically thin and all radii contribute
\citep{2011arXiv1103.0556W}.

Compared to infrared observations, there are relatively few millimeter
surveys of YSOs on account of the rapid decline of dust emission with
wavelength. The pioneering work of \cite{1990AJ.....99..924B}
showed that many protoplanetary disks in Taurus had masses comparable to or
in excess of the minimum mass solar nebula
(MMSN; $0.01\,M_\odot\simeq 10\,M_{\rm Jup}$).
This result was confirmed and extended to the similarly close
Ophiuchus star forming region by \cite{1994ApJ...420..837A}.
Improvements in detector technology led to a significantly increased
sensitivity and the ability to detect disks with masses $<1\,M_{\rm Jup}$
\citep{2005ApJ...631.1134A}.
The high resolution afforded by interferometry can distinguish
the millimeter emission from individual disks in the crowded environs
of the more distant Orion Trapezium Cluster where
\cite{2005ApJ...634..495W} and \cite{2006ApJ...641.1162E}
found several disks with the capacity to form a solar system.
In a more comprehensive survey,
\cite{2009ApJ...694L..36M} showed that the Trapezium Cluster
disk mass distribution is similar to Taurus and Ophiuchus but
truncated at the high mass end, $\gtrsim 34\,M_{\rm Jup}$,
due to photoevaporation by the $40\,M_\odot$ O6 star, $\theta^1$\,Ori\,C.

The Taurus, Ophiuchus, and Orion star forming environments are all
quite young, with typical YSO ages $\sim 1$\,Myr.
A key question is how does the disk mass evolve with time.
A handful of older, massive disks exist
\citep[e.g., TW Hydrae,][]{2000ApJ...534L.101W}
but the survey of 3\,Myr to 3\,Gyr stars by
\cite{2005AJ....129.1049C}
show that such objects are very rare.
This study of IC348 is motivated by the desire to quantify
the decline in disk masses after 1\,Myr.

IC348 is located at the edge of the Perseus molecular cloud and
has been well studied at optical to infrared wavelengths.
It contains 283
spectroscopically confirmed members in a relatively compact
($20^\prime\times 20^\prime$) region
\citep{1998ApJ...508..347L}
with a mean age estimated to lie between 2 and 3\,Myr
\citep{1998ApJ...497..736H}.
About half of the YSOs in IC348 show mid-infrared emission in
excess of the stellar photosphere \citep{2006AJ....131.1574L},
which is consistent with the median disk lifetime derived from the
statistics of many clusters \citep{2007ApJ...662.1067H}.

\cite{2002AJ....124.1593C}
mapped the central region of IC348 containing 95 YSOs at 3\,mm with the
OVRO interferometer but did not find any significant detections at a
$3\sigma$ sensitivity of $25\,M_{\rm Jup}$.
By selecting specific fields centered on concentrations of known YSOs,
rather than a contiguous mosaic, and by observing at the
significantly shorter wavelength of 1.3\,mm,
we are able to achieve an order of magnitude higher mass sensitivity than
\cite{2002AJ....124.1593C}
and make the first mass measurements of protoplanetary disks
in this region. We describe the observing strategy in \S\ref{sec:obs}
and the results in \S\ref{sec:results}. We analyze the properties of
the detected sources and compare the distribution of disk masses with
the younger Taurus, Ophiuchus, and Orion regions in \S\ref{sec:analysis},
then discuss the implications of this work for disk evolution and
planet formation in \S\ref{sec:discussion}.
We summarize our findings in \S\ref{sec:summary}.

\section{Observations}
\label{sec:obs}
\begin{figure}[tb]
\figurenum{1}
\centering
\includegraphics[width=3.5in]{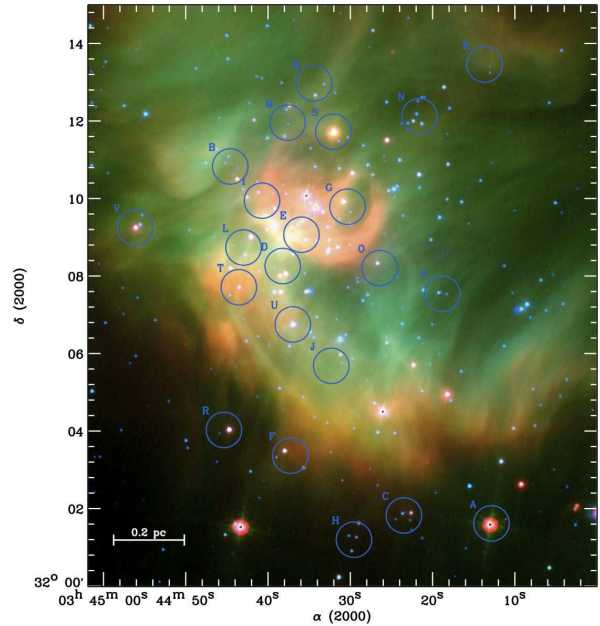}
\vskip -0.6in
\caption{Three-color infrared image of IC348 showing YSOs and nebulosity
of the associated molecular cloud.
The \emph{Spitzer} MIPS $24\,\mu$m observations are in red,
IRAC $5.8\,\mu$m in green, and IRAC $3.6\,\mu$m in blue.
The blue circles show the location and $30''$ extent of the 22 SMA fields.
The alphabetical labeling follows Table~\ref{tab:obs}.}
\label{fig:fields}
\end{figure}

We observed 22 fields containing 85 YSOs in IC348 with the
Submillimeter Array (SMA) in its compact configuration over
11 nights from October 2009 through December 2010.
The field locations were chosen to maximize the total number of
YSOs in the survey while avoiding localized regions of cloud emission
determined from $850\,\mu$m SCUBA maps
\citep{2005A&A...440..151H}.
The observed fields are overlaid on a three-color infrared image of
the region created from the
\emph{Spitzer} Heritage Archive\footnote{http://archive.spitzer.caltech.edu/}
in Figure \ref{fig:fields}.

Observations were generally carried out in average Mauna Kea
weather with precipitable water vapor levels ranging from 1.5 to 3\,mm.
IC348 transits nearly overhead in the sky providing long observational
tracks and low median system temperatures, $T_{\rm sys}=90-140$\,K.
Typically observations of 3 fields were interleaved with each
other on a single track so that the total time
per field was 2--2.5 hours, resulting in an rms noise
ranging from 0.4 to 0.9\,mJy with median 0.6\,mJy.
The field centers, dates of observation, and rms noise are
tabulated in Table~\ref{tab:obs}.

The receivers were tuned to place the 230.538\,GHz (1.3\,mm)
$J=2-1$ line of CO in the upper sideband and the same transition
of the isotopologues, $^{13}$CO, and C$^{18}$O in the lower sideband.
The line-free regions provided nearly 4\,GHz of bandwidth for
continuum measurements.

Together with the observations of IC348,
calibration measurements were made through observations of
3C454.3, 3C279, or 3C273 for bandpass,
J0336+323 and 3C84 for amplitude and phase,
and Uranus and Titan for the absolute flux scale.
The data were calibrated using standard procedures in the MIR software
package\footnote{\url{http://cfa-www.harvard.edu/$\sim$cqi/mircook.html}}.
The rms amplitude and phase noise in the calibrated data
were $\sim 20$\% and $10^\circ-20^\circ$ respectively.
The calibrated visibilities were then exported into
MIRIAD\footnote{\url{http://www.cfa.harvard.edu/sma/miriad/}}
for imaging and analysis. As protoplanetary disks are too small to be
resolved in these compact configuration data, the images were produced
using natural weighting to produce the highest signal-to-noise and an
approximately circular synthesized beam $\sim 2\farcs 5$ beam.
The flux density or upper limit toward each YSO was determined
by fitting a point source with the appropriate phase offset
in the visibility plane.

\section{Results}
\label{sec:results}
Contour maps of the 1.3\,mm emission toward each of the 22 observed
fields are in the (online) Figures~\ref{fig:mosaic1}, \ref{fig:mosaic2}.
These fields contain a total of 85 YSOs from the \cite{2003ApJ...593.1093L}
compilation, all of which were detected with \emph{Spitzer} in multiple bands
and tabulated by \cite{2006AJ....131.1574L}.
We detect 10 of these YSOs with a signal-to-noise ratio greater
than 3. These are labeled in the figures with the identification
number from Table~1 of \cite{2006AJ....131.1574L} for ease of reference
to the mid-infrared fluxes later.

There are also a small number of possibly significant emission
peaks that do not correspond to the position of a known YSO that may
be due to fluctuations in the background cloud emission
or sidelobes from strong emission outside of the primary beam
\citep{2005A&A...440..151H}.
Strong CO and, in many cases, isotopologue emission is also found
but the images do not show clear evidence for point source emission
toward the YSOs and we are unable to constrain the gas content of
the disks.  Cloud contamination is much more severe in the line
observations, possibly due to the emission being optically thick and perhaps
more spatially variable than the continuum.  A similar situation was
found in the Trapezium Cluster \citep{2010ApJ...725..430M}.

The strongest source, by far, is a core that is apparent as an
isolated point source in the SCUBA map of \cite{2005A&A...440..151H}.
This is clearly a less evolved,
protostellar source and we describe it briefly in the following subssection
for completeness. As the emission in this source is dominated by the
protostellar envelope,
we do not include it in the discussion and analysis of the protoplanetary
disks thereafter.

\subsection{The Class I Protostar, IRAS 03410+3152}
This bright object was chosen to lie in the first field as a simple
test of the observational setup.
It is a well known Class I YSO,
cataloged as IRAS 03410+3152,
source 51 in \cite{2006AJ....131.1574L},
source 101 in \cite{2005A&A...440..151H},
and source 29 in the \emph{Chandra} X-ray study of the cluster by
\cite{2002AJ....123.1613P}.
It has a bolometric temperature of 463\,K and luminosity of $1.6\,L_\odot$.
\citet{2006AJ....132..467W} also detect several $H_{2}$ shocks on
either side of the star.

\begin{figure}[tb]
\figurenum{3}
\centering
\includegraphics[width=3.4in]{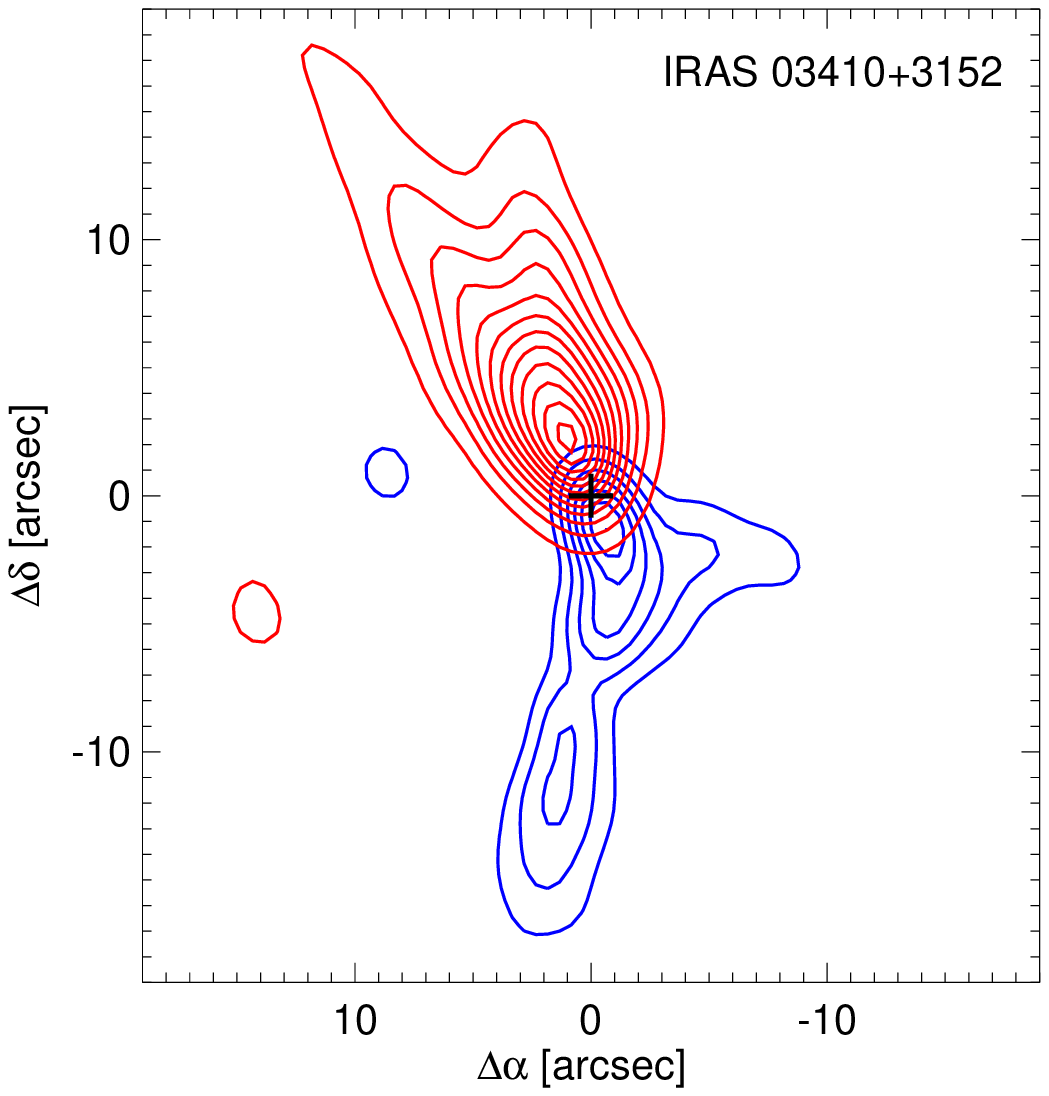}
\caption{The $^{12}{\rm CO}(2-1)$ outflow in the Class I source,
IRAS 03410+3152.
The blue contours show the integrated intensity
over the velocity range $4.5 - 6$\,km\,s$^{-1}$,
while the red contours show the intensity integrated over
$11 - 19$\,km\,s$^{-1}$.
The black cross marks the position of the continuum emission peak.
Red contour levels start at 0.8 Jy and increase by
multiples of 2.4 Jy, while blue contour levels start at 0.2 Jy
and increase by multiples of 0.2 Jy.}
\label{fig:outflow}
\end{figure}

We detect both strong continuum and line emission toward
IRAS 03410+3152. The continuum is a point source at the resolution
of our data (Figure~\ref{fig:mosaic1}) but the CO 2--1 emission shows
prominent red-blue outflow lobes with a moderate opening angle
(Figure~\ref{fig:outflow})
characteristic of Class I sources \citep{2006ApJ...646.1070A}.

IC348 abuts an active star forming region in the Perseus cloud.
IRAS 03410+3152 belongs to this younger population but the SMA field
contained a second source (Lada source ID 468)
that is a bona-fide Class II YSO.
Our detection of this object in the same field provides a graphic
illustration of the tremendous difference in the millimeter flux
from a envelope-dominated to disk-dominated YSO.
The rest of this paper concerns the nature of the latter objects.

\subsection{Protoplanetary Disk Masses}
\label{sec:masses}
Nine of the remaining 84 YSOs in the 22 SMA fields were detected with
1.3\,mm fluxes ranging from 2 to 6.5\,mJy.  The $3\sigma$ limits
on the fluxes of the 75 non-detections were typically less than 2\,mJy.
We stacked the emission of these non-detections to constrain their
average properties. There was no clear detection in the stacked map
with a $3\sigma$ limit to the average flux of 0.3\,mJy.

As the millimeter emission is optically thin, it provides a direct
measure of the dust mass. We convert fluxes to total disks masses
through the simple formula,
$$
M_{\rm d} = {F_\nu d^2\over\kappa_\nu B_\nu(T)},
$$
where the dust opacity,
$\kappa_{230{\rm GHz}}=0.023\,{\rm cm}^2\,{\rm g}^{-1}$,
follows the commonly used prescription from \cite{1990AJ.....99..924B}
and implicitly assumes an ISM gas-to-dust ratio of 100.
We use a distance $d=320$\,pc based on \cite{1998ApJ...497..736H}.
Although the dust in the disks lies over a range of radii and therefore
temperatures, the Planck function, $B_\nu$, is calculated at a
a single temperature 20\,K based on the modeling by \cite{2005ApJ...631.1134A}
who showed this provides an excellent fit to detailed modeling of
the infrared-millimeter spectral energy distribution (SED).

The vernacular of disk mass, as derived in this way,
is standard in the literature but is really a shorthand for a
procedure that has much more validity
in the interstellar, as opposed to circumstellar, medium.
It is important to keep in mind that we are really only estimating
the mass of micron to millimeter sized particles and then extrapolating
the gas content by multiplying by two orders of magnitude!

The inferred disk masses vary between 2 and $6\,M_{\rm Jup}$.
Their properties are tabulated in Table~\ref{tab:det}.
The survey is 100\% complete for disk masses
$M_{\rm d}>2.3\,M_{\rm Jup}$.
The limit on the stacking of the non-detections implies that the
average mass of the non-detections,
$\langle M_{\rm d}\rangle < 0.27\,M_{\rm Jup}$.

The nine detections, although only a small fraction of the entire sample,
provide the first disk mass measurements in IC348.
They provide important constraints on the disk evolution at 2--3\,Myr
with commensurate implications for understanding planet formation.
We place these measurements in context with disk masses in younger
star forming regions below after first comparing the properties of
the detected and non-detected sources.

\section{Analysis}
\label{sec:analysis}
\subsection{Comparison of Detected and Non-detected Sources}

\begin{figure}[tb]
\figurenum{4}
\centering
\includegraphics[width=3.4in]{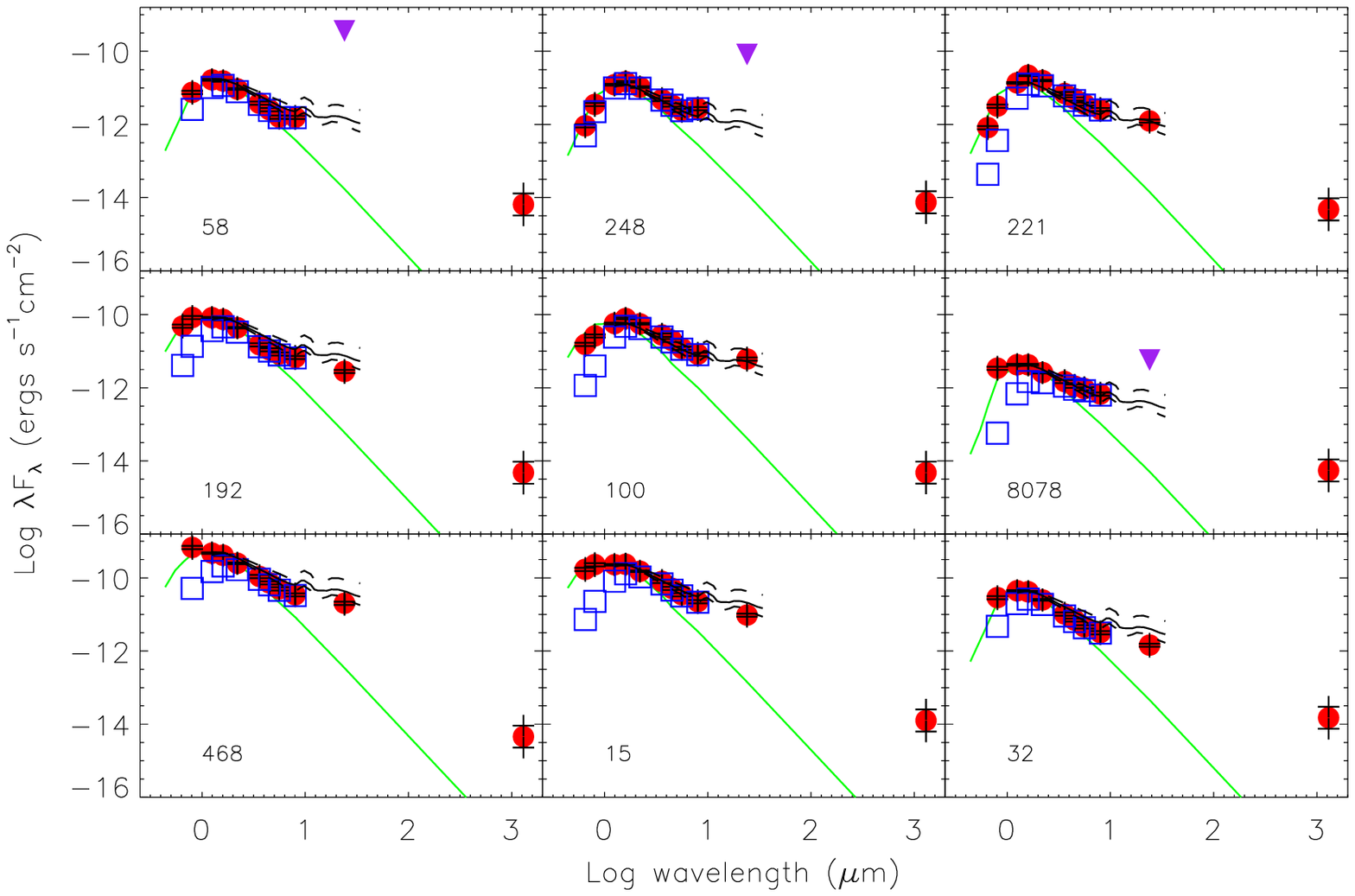}
\caption{Spectral Energy Distributions for the nine SMA detected disks
labeled by the IDs in Table~\ref{tab:det}.
The blue squares represent the observed optical and IRAC fluxes,
uncorrected for extinction, while the filled red circles represent
the extinction-corrected optical and IRAC fluxes, as well as the
MIPS $24\,\mu$m and SMA $1.3\,$mm fluxes. 
Purple diamonds represent upper limits to the MIPS $24\,\mu$m flux,
and the black crosses represent 1-$\sigma$ error bars.
The green lines show model photospheres for the published
spectral types of each star, and the black solid and dashed lines
show the median and quartiles, respectively, of the infrared
SEDs for CTTS, scaled to the $J$-band flux.}
\label{fig:sed}
\end{figure}

Figure~\ref{fig:sed} plots the SED of the nine detected sources.
The mid-infrared fluxes in \emph{Spitzer} IRAC bands 1--4
($3.6-8.0\,\mu$m) and MIPS band 1 ($24\,\mu$m) are from
\cite{2006AJ....131.1574L}.
Optical and near-infrared flux densities in the
$R_{\rm C}$, $I_{\rm C}$, $J$, $H$, and $K_{\rm s}$-bands were obtained by
cross-referencing the source positions in \cite{2006AJ....131.1574L}
with tables published by \cite{2007ApJ...667..308C}
and \cite{2003ApJ...593.1093L}.
We estimate the extinction,
$A_{\rm V} = 4.76E(R-I)
           = 4.76[(R_{\rm C}-I_{\rm C})_{\rm obs} - (R_{\rm C}-I_{\rm C})_{0}]$,
as in \cite{2007ApJ...667..308C}
where the intrinsic stellar colors, $(R_{\rm C}-I_{\rm C})_{0}$,
are from \cite{1995ApJS..101..117K}
for the spectral types as determined by \cite{2003ApJ...593.1093L}.
The infrared SEDs are similar, but marginally lower in a handful of cases,
than the median SED of classical T-Tauri stars (CTTS)
out to $24\,\mu$m, which is over-plotted in each panel.

\begin{figure}[tb]
\figurenum{5}
\centering
\includegraphics[width=3.5in]{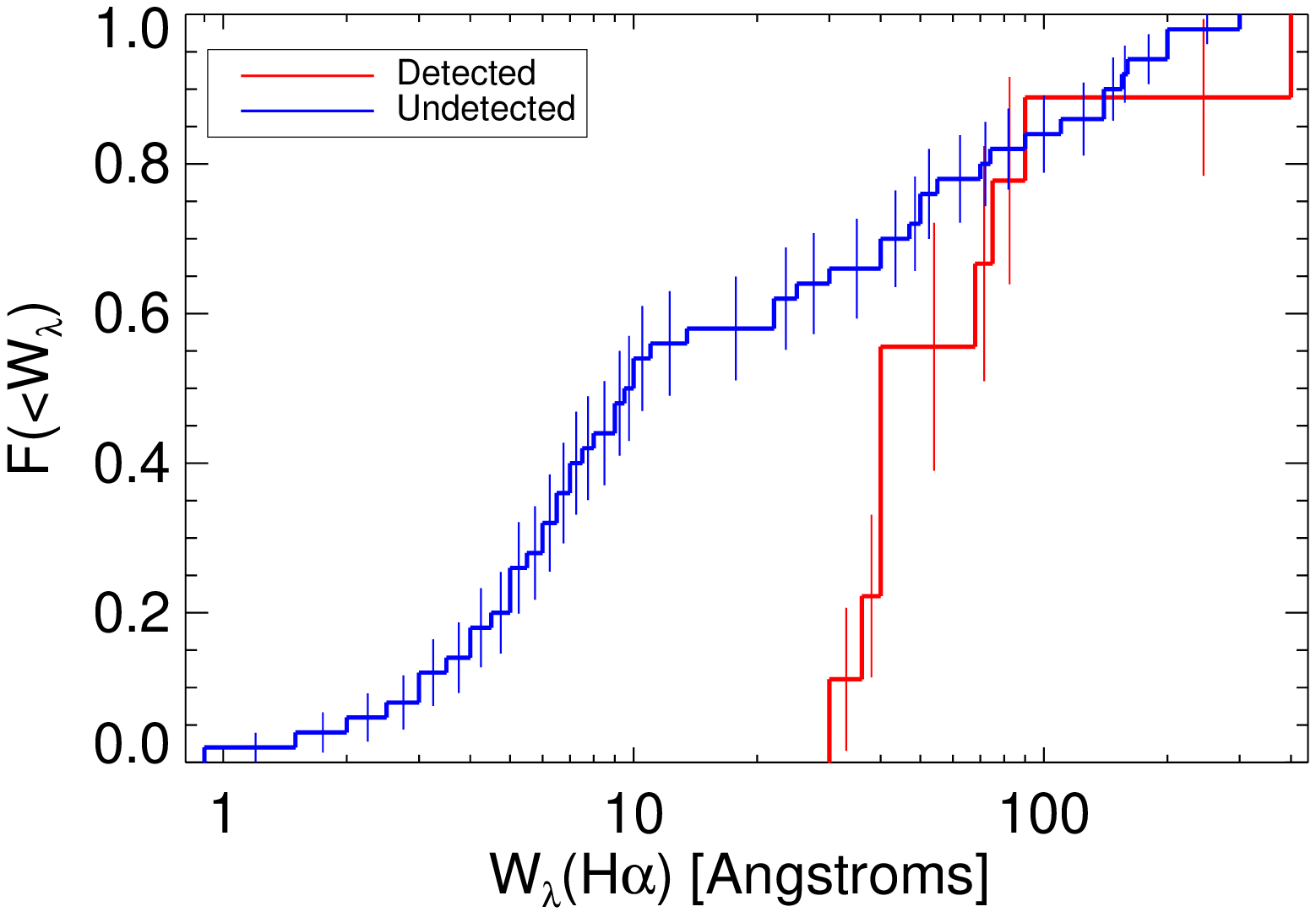}
\caption{Cumulative distributions of H$\alpha$ equivalent widths for
the SMA detected (red line) and non-detected sources (blue line),
for the 60 out of 84 surveyed disks where such measurements exist.}
\label{fig:EW}
\end{figure}

Many of the millimeter non-detections have quite similar infrared SEDs
as the millimeter detections in Figure~\ref{fig:sed}.
However, there is a statistical difference in the accretion
properties of the two samples.
Figure~\ref{fig:EW} plots the cumulative distribution of the
H$\alpha$ equivalent widths, $W_\lambda({\rm H}\alpha)$,
for 60 of the 84 disks in our survey from
\cite{2003ApJ...593.1093L}.
All nine disks detected at 1.3\,mm have large $W_\lambda({\rm H}\alpha)$,
indicative of high accretion, but most of the non-detections have
much lower values.  About two-thirds of the non-detections with
$W_\lambda({\rm H}\alpha)$ measurements have equivalent widths
less than the minimum of the detected disks.
Based on the spectral type dependent $W_\lambda({\rm H}\alpha)$ cutoff
defined by \cite{2003ApJ...582.1109W},
most of the sources that are not detected in our SMA survey
are weak-lined T-Tauri stars (WTTS).
However, we only detect 9 out of the 25 identified CTTS.
The sensitivity limitations of our survey prevent us from drawing
any strong conclusions, but the general correspondence between the
millimeter detections and high $W_\lambda({\rm H}\alpha)$
suggests that the gas and dust depletion timescales are not
greatly dissimilar.

\begin{figure}[tb]
\centering
\figurenum{6}
\includegraphics[width=3.0in]{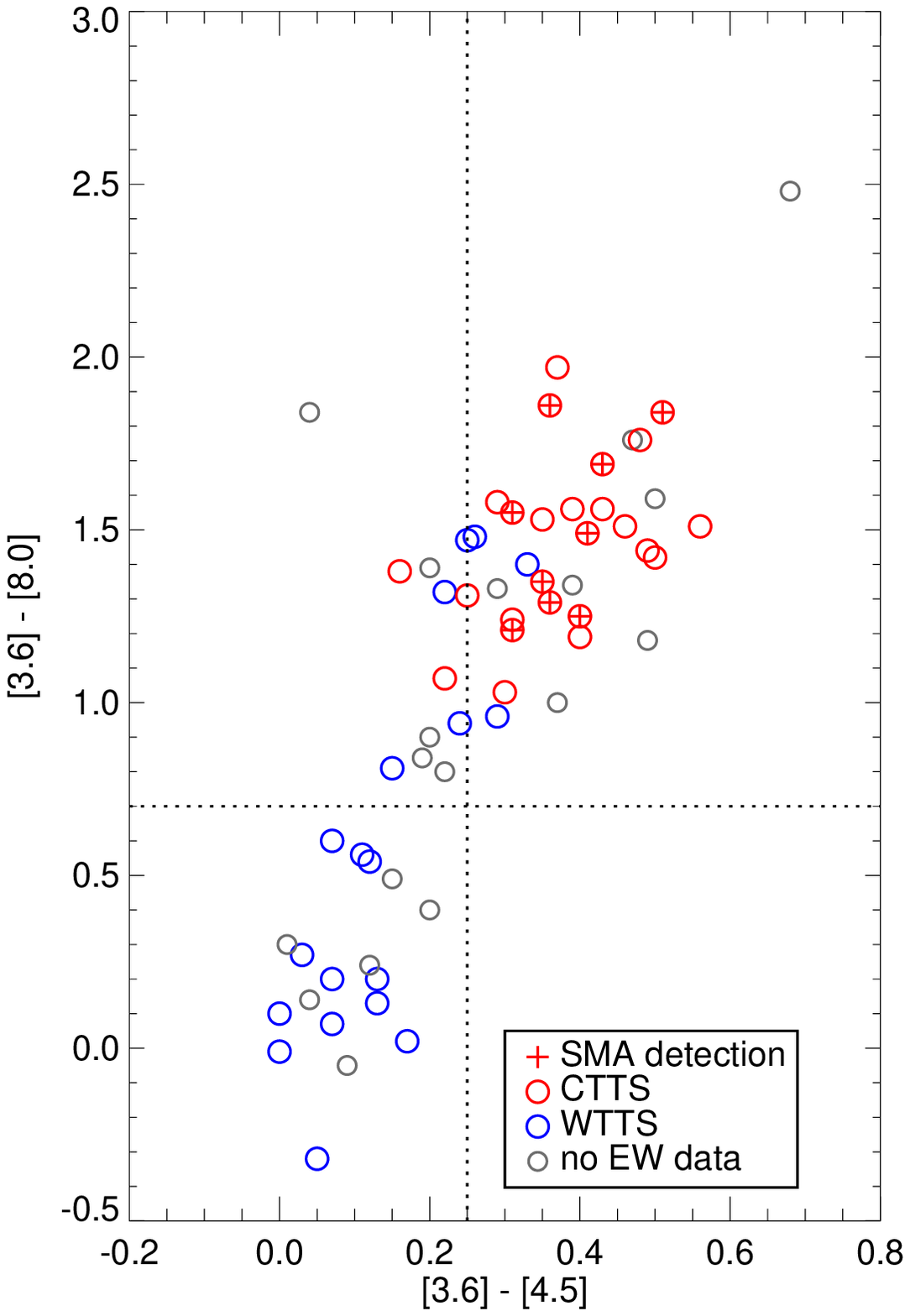}
\caption{Infrared color-color plot of the surveyed YSOs
from \emph{Spitzer} IRAC bands 1, 2, and 4.
The abscissa is the magnitude difference between 4.5 and $3.6\,\mu$m
and the ordinate is the magnitude difference between 8.0 and $3.6\,\mu$m.
The dashed lines show the boundaries where there a clear infrared excess
above the stellar photosphere.
The symbols characterize the sources through their H$\alpha$
equivalent width and millimeter emission.
The SMA detections are all accreting and have infrared excesses
at both 4.5 and $8.0\,\mu$m.}
\label{fig:colcol}
\end{figure}

The situation is encapsulated in the IRAC color-color plot in
Figure~\ref{fig:colcol} where different symbols
represent CTTS, WTTS and SMA detections.
The vertical and horizontal dotted lines show the boundaries used by
\cite{2007ApJ...667..308C}
to distinguish between YSOs with colors consistent with a stellar
photosphere (lower left corner), transition disks with excess
infrared emission only apparent at wavelengths greater than $4.5\,\mu$m,
and ``full'' disks with excesses at $4.5$ and $8\,\mu$m.
All the CTTSs in our sample display excess emission at $8\,\mu$m,
and the nine disks detected by the SMA exhibit excess emission at
both $4.5\,\mu$m and $8\,\mu$m.  However, most disks with an infrared excess
were not detected in our millimeter survey and there is no correlation
between disk mass and the amount of infrared excess.
We stacked the SMA emission from all of the stars with infrared excesses
that were not individually detected with the SMA, but did not find a
significant detection.

\begin{figure*}[!ht]
\figurenum{2a [Online-only]}
\centering
\includegraphics[width=6.5in]{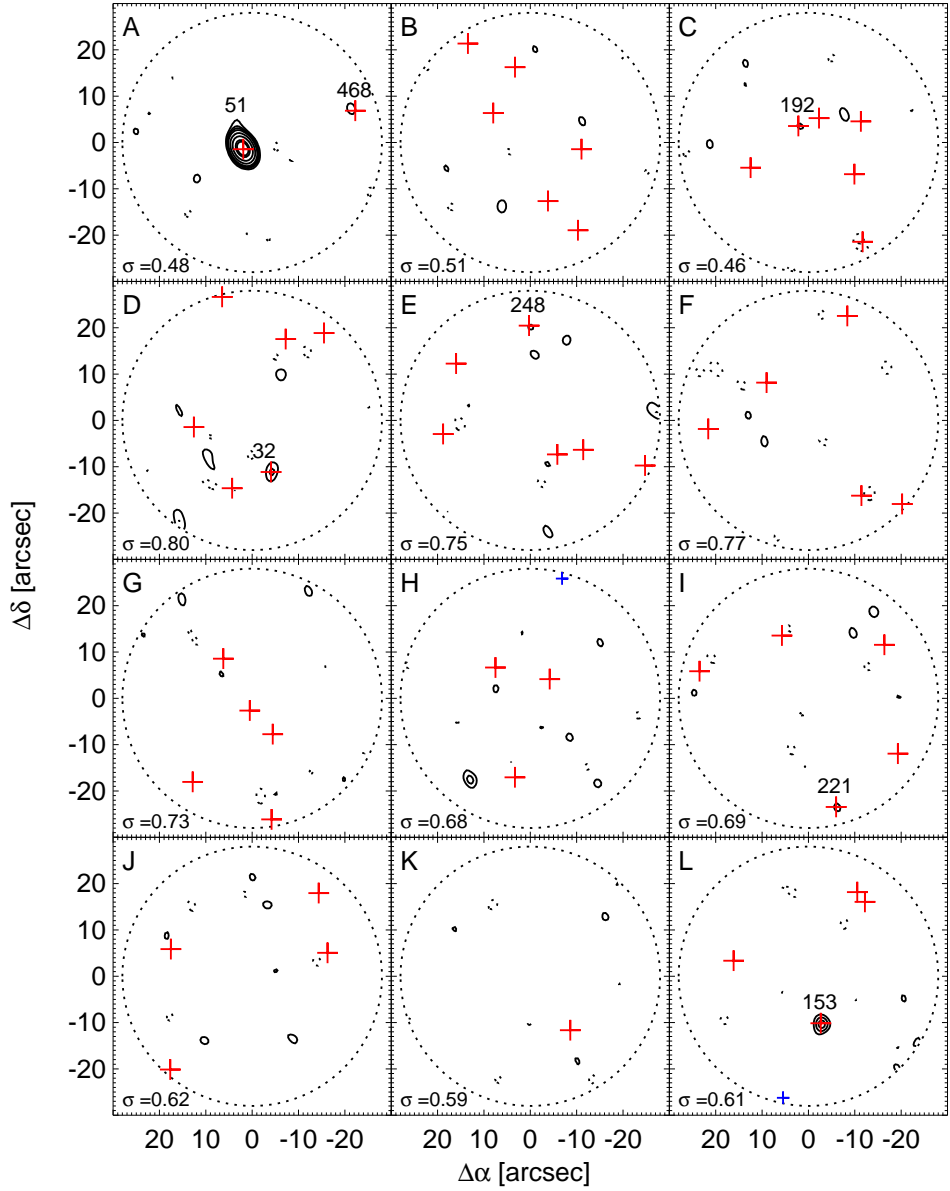}
\caption{Cleaned SMA maps of the 22 fields in the survey.
The letter at the top left of each panel corresponds to the
labels in Figure~\ref{fig:fields}.  The rms noise, $\sigma$ (mJy),
in each image is given in the bottom left of each panel.
Contour levels begin at $3\sigma$ and increase in steps of $2\sigma$.
Dotted contours show negative emission on the same scale.
The locations of the cluster members from the
\cite{2003ApJ...593.1093L} compilation are shown by red crosses.
All these sources were also detected with \emph{Spitzer}
by \cite{2006AJ....131.1574L}.
There are an additional 5 sources with \emph{Chandra} X-ray detections
in the \cite{2002AJ....123.1613P} compilation.
These are most likely extragalactic sources and one,
PZ195 in field V, shows significant 1.3\,mm emission.
The SMA detected YSOs are labeled by their ID number in
Table~\ref{tab:det}.}
\label{fig:mosaic1}
\end{figure*}

\begin{figure*}[!ht]
\figurenum{2b}
\centering
\includegraphics[width=6.5in]{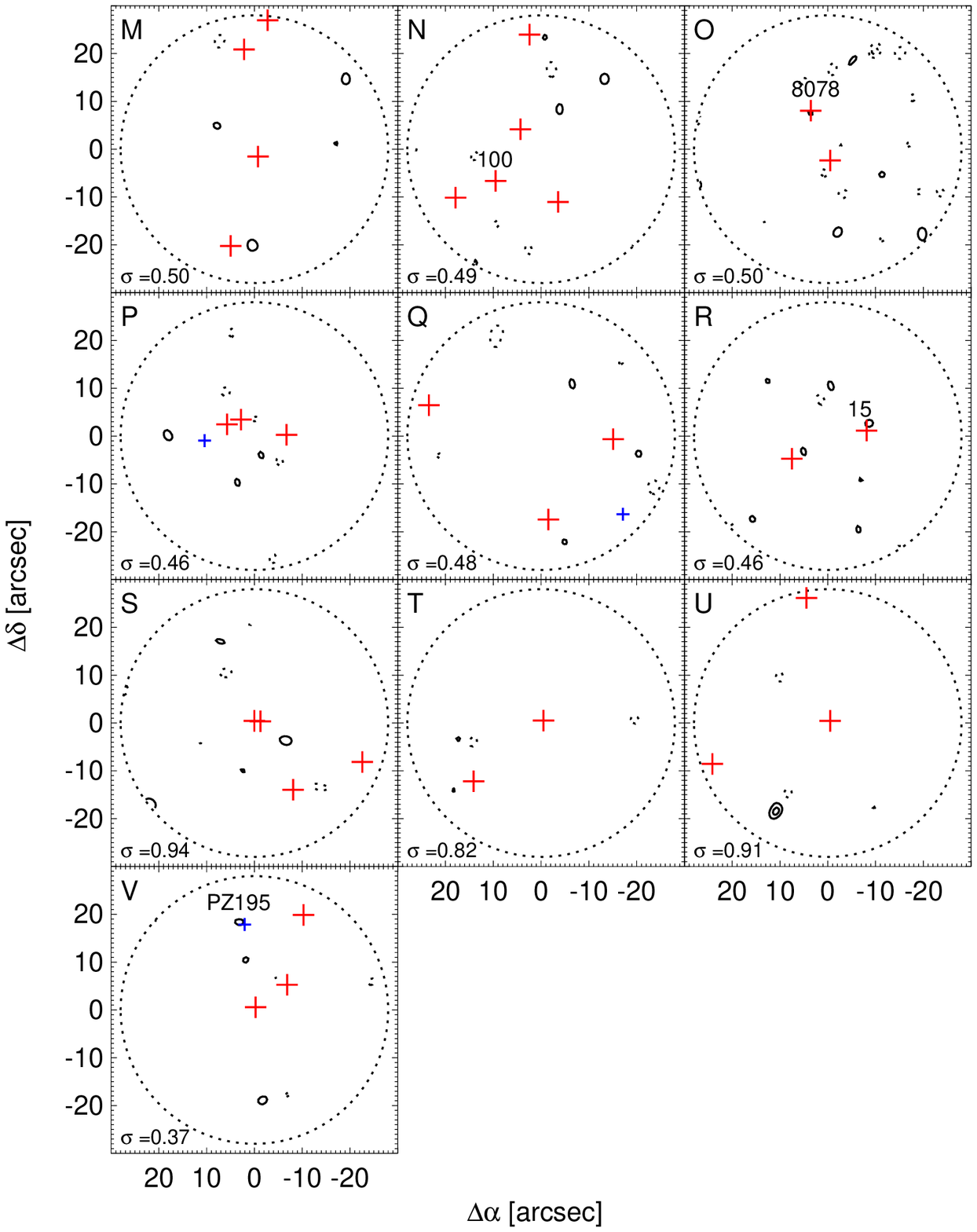}
\caption{Fig 2 (contd.)}
\label{fig:mosaic2}
\end{figure*}

There are 45 \emph{Chandra} X-ray sources within our survey
\citep{2002AJ....123.1613P}.
40 of these are coincident with known cluster members from
\cite{2003ApJ...593.1093L} and are active YSOs.
5 were detected with the SMA (including IRAS 03410+3152)
but we do not see any clear trend between the millimeter
properties of the X-ray detected and undetected sources.
The other 5 \emph{Chandra} sources have no optical or infrared counterpart
to a level that rules out highly extincted substellar objects.
These are indicated by blue crosses in
Figures~\ref{fig:mosaic1},\ref{fig:mosaic2} and,
as \cite{2002AJ....123.1613P} concluded,
are most likely extragalactic contaminants.
Interestingly, we detect one of these, source 195
in field V (the lowest noise map in our survey),
with a flux $1.2\pm 0.4$\,mJy.

Most YSOs in IC348 are between 2 and 3\,Myr old
but there is likely to be a significant dispersion.
A simple explanation of our results would be that the detected
disks are younger than the non-detections.
In principle, protostellar ages and masses can
be inferred by comparing their luminosity and effective temperature
against model isochrones of pre-main-sequence tracks but,
in practice, the associated errors are large.
Using the model isochrones of \cite{1997MmSAI..68..807D},
we do not find any significant difference in the estimated
ages of the millimeter detected and non-detected YSOs.
This is not too surprising given previous failed searches for differences
between CTTS and WTTS in individual clusters \citep{1992AJ....104..762G}
which show that age is not the sole determinant of disk evolution.

Protoplanetary disk masses are also known to scale with stellar mass
\citep{2011arXiv1103.0556W}.
The inferred stellar masses of the detected disks are all sub-solar
but range widely from 0.02 to $0.77\,M_\odot$. In this small sample,
there is no evidence for a dependency on stellar mass but the detection
of a disk around source 468, a M8.25 brown dwarf, is noteworthy.

\subsection{Comparison with Taurus, Ophiuchus, and Orion}
\begin{figure}[tb]
\centering
\figurenum{7}
\includegraphics[width=3.56in]{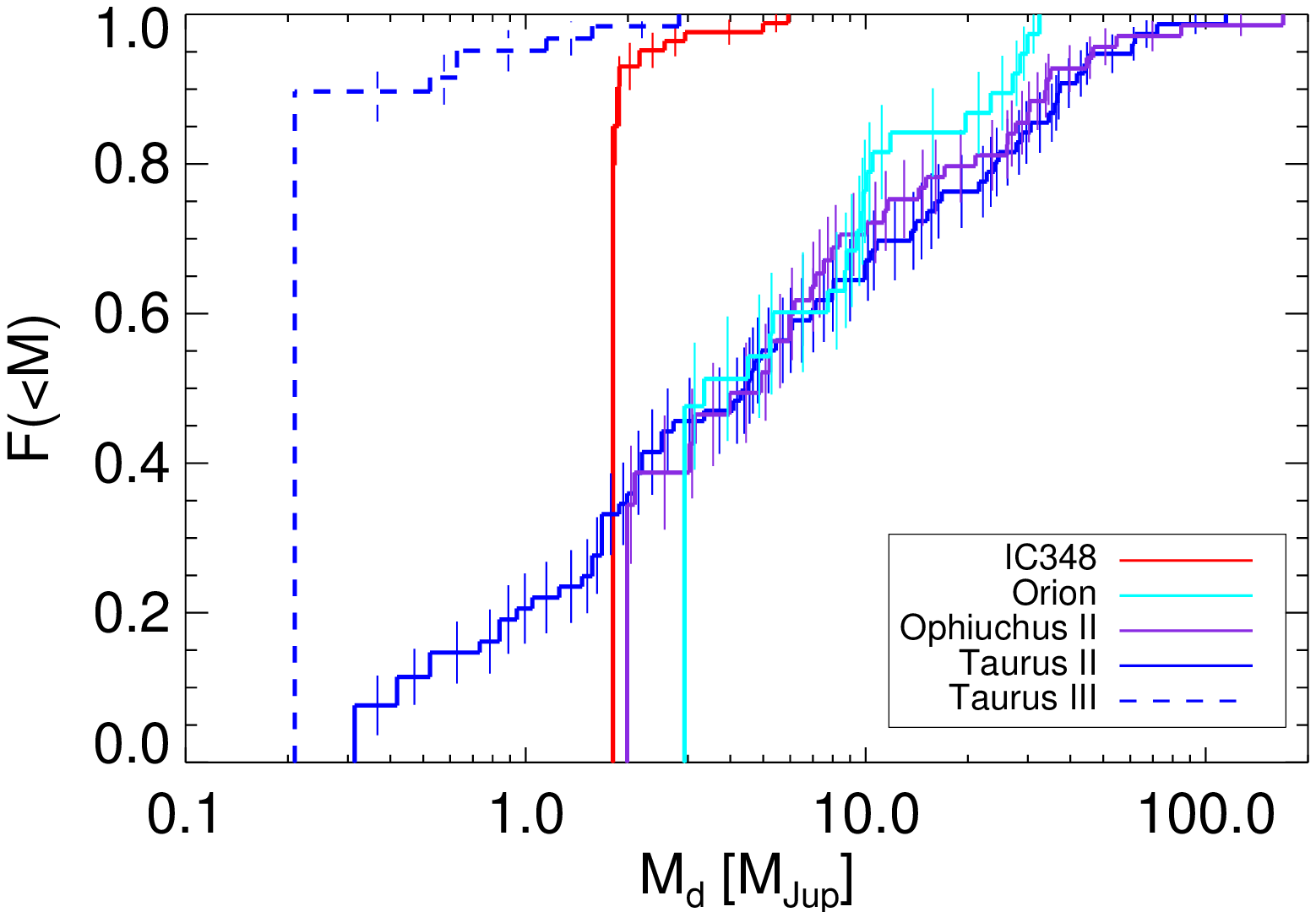}
\caption{Cumulative distributions of protoplanetary disk masses
in IC348, Taurus, Ophiuchus, and Orion.
The mass distribution for non-accreting Class III sources in Taurus
is shown as the blue dashed line.
The distributions were calculated using a Kaplan-Meier product limit
estimator that allows the incorporation of 3$\sigma$ upper limits into
the distributions. The long vertical lines dropping down to zero
indicate the lowest mass or upper limit in the sample.} 
\label{fig:mass}
\end{figure}

Based on the lack of detections of relatively massive disks,
\cite{2002AJ....124.1593C} concluded that disks
in IC348 have significantly lower masses than those in Taurus.
The high sensitivity of this SMA survey has allowed us to make the
first mass measurements of IC348 disks and the results provide
an important benchmark for tracking disk evolution at timescales
comparable to their statistical half-life.
Due to the large number of non-detections and because the sensitivity
of the observations varied from field to field,
we plot the cumulative disk mass distribution,
$F(<m)$, in Figure~\ref{fig:mass}
using the Kaplan-Meier product limit estimator
\citep{1985ApJ...293..192F}.
Similarly defined distributions are plotted for the (sub-)millimeter
derived disk mass measurements in Taurus \citep{2005ApJ...631.1134A},
Ophiuchus \citep{2007ApJ...671.1800A},
and Orion \citep{2010ApJ...725..430M}.

\begin{figure}[tb]
\centering
\figurenum{8}
\includegraphics[width=3.5in]{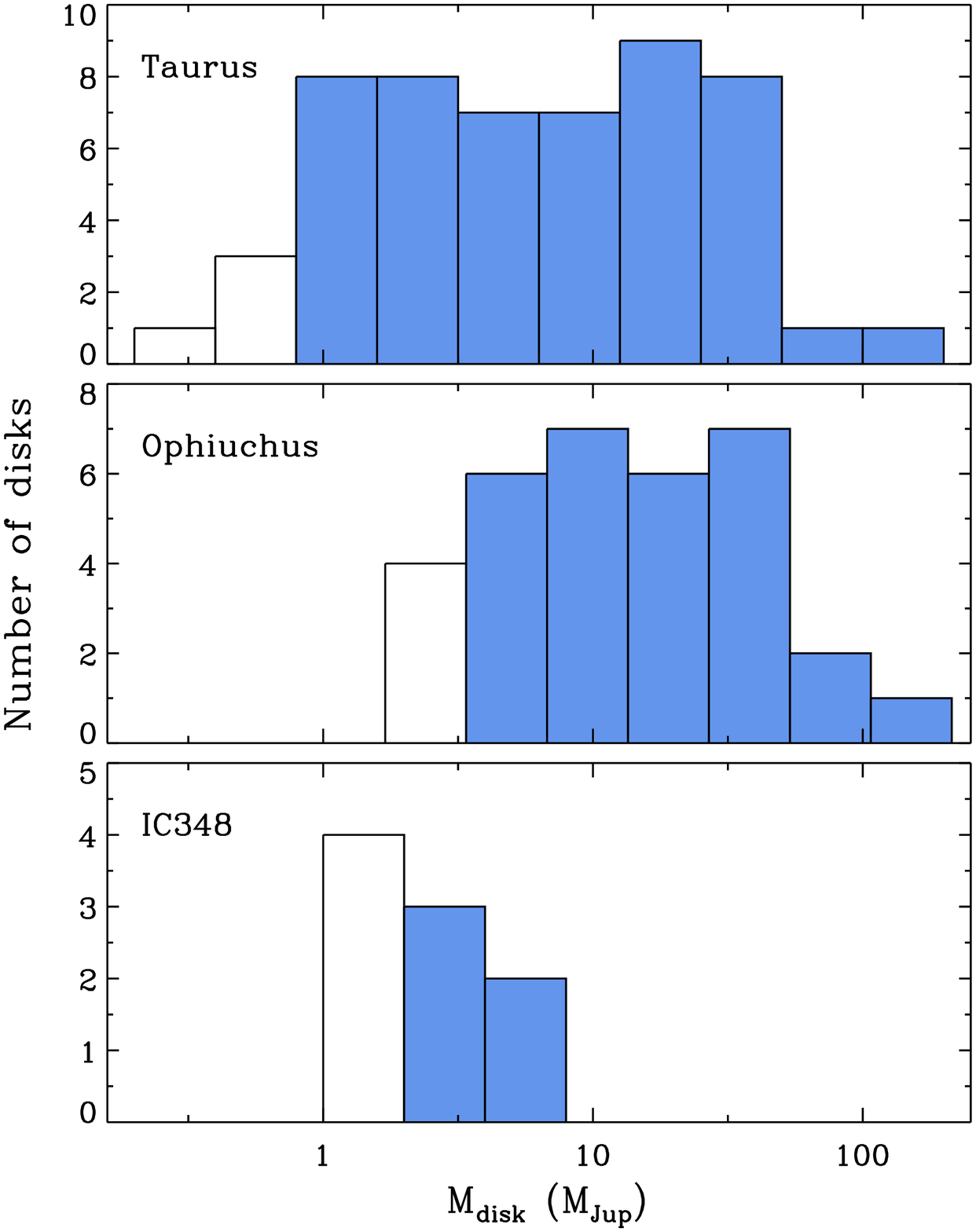}
\caption{Mass histogram of protoplanetary disk masses
in Taurus, Ophiuchus, and IC348.
The blue bins show data for which the survey in each region is complete.}
\label{fig:masshist}
\end{figure}

As noted by \cite{2007ApJ...671.1800A},
the Taurus and Ophiuchus distributions are remarkably similar.
Because of the greater distance to Orion, the completeness
limit is higher but also shows a similar distribution between
$\sim 3-10\,M_{\rm Jup}$ with a decline at higher masses
due to photoevaporation \citep{2010ApJ...725..430M}.
A more conventional, binned distribution of the Taurus, Ophiuchus
and IC348 disk masses is plotted in Figure~\ref{fig:masshist}.

The disk mass distribution in IC348 is obviously very different
from these three younger star-forming regions. Our inferred
masses, which are the high mass tail of the full distribution,
are comparable to the median mass in the other regions.
Or, to put it another way, more than 60\% of Taurus and Ophiuchus
disks, but less than 10\% of IC348 disks, have masses greater
than $2\,M_{\rm Jup}$.
Further, although the number of detected disks is small,
there is no sharp falloff at the high mass end of the IC348 distribution
as in Orion. Rather, the IC348 disk mass distribution seems
to be systematically shifted down by about a factor of 20
relative to Taurus and Ophiuchus,
$$F_{\rm IC348}(<m)\simeq F_{\rm Taurus}(<20m),~~~~~m>2\,M_{\rm Jup}$$

The comparison here is of Class II YSOs, or optically visible 
stars with infrared excesses and signatures of accretion but no
envelope material. A handful of Class III YSOs (no infrared excess
or accretion) were detected in the sensitive Taurus survey by
\cite{2005ApJ...631.1134A}
and their distribution is plotted
as the dashed line in Figure~\ref{fig:mass}.
The 2--3\,Myr IC348 Class II disks
are more massive than the $\sim 1$\,Myr Taurus Class III sources.
We conclude that we are witnessing the tail end of the Class II phase,
just before accretion ends and the disk disappears.

\cite{2003ApJ...593.1093L} find that the distribution of stellar
masses peaks at lower masses in IC348 compared to Taurus.
The difference between the stellar mass functions is most pronounced
above $1\,M_\odot$. Almost all the stars used for the disk mass function
comparison here are spectral types K and M and have sub-solar masses.
We also compared the disk mass distribution for M stars only and find,
with statistically less precision, the same large difference.
The dramatically lower disk masses in IC348 are not due to differences
in stellar mass.

Because of the steps involved in determining masses from
millimeter fluxes, there may be systematic effects in the
comparisons between different regions.
In each case, however, the masses were determined assuming the same
characteristic dust temperature of 20\,K, 
the same \cite{1990AJ.....99..924B} dust opacity,
$\kappa_\nu=0.1(\nu/10^{12}\,{\rm Hz})\,{\rm cm^2\,g}^{-1}$,
and the same gas-to-dust ratio of 100.
The average dust temperature and frequency dependence of dust opacity
can be checked with SED modeling and are likely to be reasonably accurate.
Cloud confusion prevents us from making an unambiguous CO measurement
of gas in the disks but using the same gas-to-dust ratio for all regions
allows the simplest comparison.
Taking this factor out, we can state with more certainty that
there is a reduction of about a factor of 20 in the amount of
millimeter and smaller sized dust grains in the IC348 disks
compared to the younger regions and, presumably, this is
due to the additional $\sim 1-2$\,Myr of evolution.

\section{Discussion}
\label{sec:discussion}
Sensitive infrared surveys of many clusters of different ages have
firmly established the trend of decreasing disk fraction with age
and a median disk lifetime around low mass stars of $\sim 2-3$\,Myr
\citep{2007ApJ...662.1067H}.
The fraction of accreting systems declines on a similar,
though possibly slightly shorter timescale
\citep{2010A&A...510A..72F}.
IC348 disks are old enough that we expect many to be fully
depleted of both dust and gas but we have been able to quantify
the amount of dust in a few of the sources with infrared excesses
and found that they are all active accretors with H$\alpha$
equivalent widths greater than the cluster average.

There is a wide diversity of disk SEDs in IC348
\citep{2006AJ....131.1574L, 2009AJ....138..703C}.
Based on \emph{Spitzer} IRAC and MIPS data, they consider three categorizations:
(1) anemic disks with excess emission above the stellar photosphere
in all IRAC bands ($\lambda=3.6-8\,\mu$m)
but at levels significantly below the median disk SED in Taurus;
(2) transition disks or disks with inner holes
that have a mid-infrared dip in their SED, with strong excess emission
only at $\lambda\geq 24\,\mu$m;
(3) primordial or ``full'' disks that show excess emission at all bands
at levels similar to the median disk SED in Taurus.
At our survey detection threshold of about 2\,mJy,
corresponding to $2\,M_{\rm Jup}$ under the assumptions in
\S\ref{sec:masses}, we only detected a handful of primordial disks.
The others, including most primordial disks and all anemic and
transition disks, have very low masses.

\cite{2006AJ....131.1574L} suggested that the low infrared excesses in
anemic disks are due to a flattened geometry arising from dust settling.
Our observations show that there is a paucity of millimeter-sized dust
grains, however, and the disk has not just settled but has largely depleted.
Similarly the low masses of the transition disks reveals that, in general,
the inner disk only dissipates after the outer disk is largely depleted.
\cite{2008ApJ...686L.115C} came to the same conclusion based
on their low limits to the disk masses of WTTS.
This is consistent with photoevaporative models of disk evolution
whereby the outer disk feeds the inner disk until the viscous accretion
rate drops below the photoevaporation rate from the central star.
Our low disk mass upper limits suggest that the
photoevaporation rate is low, consistent with FUV radiation
\citep[and references therein]{2011arXiv1103.0556W}.
In fact, the disks that we have detected are the least evolved,
in the sense that they exhibit strong infrared excesses at all wavelengths
and H$\alpha$ emission indicative of gas accretion.
Yet we only detect about one third of such CTTS and their masses
are all substantially below the canonical $10\,M_{\rm Jup}$ MMSN.

Assuming that the giant planet frequency in IC348 is to match that
measured in the field, the absence of MMSN disks implies that planet
formation is either complete or has progressed substantially.
However, the possibility that giant planets have already formed
is hard to reconcile with core accretion models 
\citep{1996Icar..124...62P, 2005Icar..179..415H}.
The formation of Jupiter and, especially, Saturn ($0.3\,M_{\rm Jup}$)
mass planets requires several Myr, significantly longer than
the median YSO age in IC348.

On the one hand, our results may be considered a point in favor of
very rapid giant planet formation through gravitational instability
\citep{1997Sci...276.1836B}.
On the other hand, if planets form over several Myr through core
accretion, this process must have progressed to the point where most of the
solid material is in large bodies and beyond our ability to detect directly
\citep{2010MNRAS.407.1981G}.
At our observing wavelength, $\lambda=1.3$\,mm, the survey is effectively
blind to particles with sizes $a\gtrsim 3\lambda \simeq 4$\,mm
\citep{2006ApJ...636.1114D}.
For a constant dust mass, and a grain size distribution, $n(a)\sim a^{-3.5}$,
the millimeter emission drops by our measured factor of 20 once the maximum
grain size reaches about a meter, as verified in the detailed models of
\cite{2001ApJ...553..321D}.


We suggest that the primordial disks with millimeter emission in IC348
may be excellent candidates for studying the progression from planetesimals
to planets.
These are the most massive disks in the young cluster but,
given their low millimeter luminosities, most of the dust
mass probably resides in macroscopic objects.
Their continuous infrared SEDs show that no planets have dynamically
cleared out radial gaps or holes that are optically thin to starlight
and their large H$\alpha$ luminosities indicate that they retain some gas.
Recent calculations by \cite{2010MNRAS.404..475J} and \cite{2010A&A...520A..43O}
show that centimeter-sized ``pebbles'' in gas rich disks experience strong
drag forces and rapidly accrete
onto protoplanets thereby accelerating planetary growth.
Our conclusion that most of the solids in the disks have aggregated
beyond millimeter sizes within 2--3\,Myr is also
consistent with cosmochemical measurements of the absolute ages
of chondrules within rocky meteorites \citep{2002Sci...297.1678A}.


\section{Summary}
\label{sec:summary}
We have conducted a 1.3\,mm SMA survey of 85 YSOs in the nearby
cluster IC348.  With a mean age of 2--3\,Myr,
the cluster presents an opportunity to study YSOs at a critical time
in their evolution when they begin to lose a large fraction of their disks.
Our main findings are:

\begin{itemize}
\item Nine protoplanetary disks are detected with inferred masses
ranging from $2-6\,M_{\rm Jup}$.

\item The millimeter detected disks are all actively accreting and
exhibit excess photospheric emission above the photosphere at
$\lambda\geq 4.5\,\mu$m with infrared SEDs similar to the median in Taurus.

\item The protoplanetary disk mass distribution in IC348 is shifted
to lower masses by about a factor of 20 compared to the
$\sim 1$\,Myr old Taurus and Ophiuchus star-forming regions.
We suggest that this is due to substantial grain growth such
that most of the solid mass resides in large particles,
$\gg 4$\,mm and perhaps up to meters, in at least some
of the disks.

\item Grain growth beyond millimeter sizes in gaseous disks may
promote rapid planetary growth and the disks that we have detected
with the SMA likely signpost the best candidates in the cluster for
witnessing the birth pangs of giant planets.

\end{itemize}

Future directions include resolved observations to see whether any
protoplanets might have opened up ``dusty gaps'' that are 
infrared bright but millimeter-faint \citep{2011ApJ...732...42A}.
It will also be very interesting to try additional spectroscopic
observations to measure their gas content, and to observe them
at longer wavelengths to constrain the population of centimeter
and larger sized grains. Due to their low millimeter luminosities,
however, these are tasks for upcoming facilities.

\acknowledgements
We thank Anders Johansen, Jes Jorgensen, and Sean Andrews for
interesting discussions on the topics of grain growth and disk evolution.
This work is supported by the NSF through grant AST08-08144.

\clearpage
\begin{deluxetable}{ccccc}
\tablecaption{Observed fields\label{tab:obs}}
\tablenum{1}
\tablewidth{0pt}
\tablehead{\colhead{Field} & \colhead{R.A.} & \colhead{Decl} & \colhead{Date} & \colhead{RMS} \\
\colhead{} & \colhead{(J2000)} & \colhead{(J2000)} & \colhead{(yymmdd)} & \colhead{(mJy)}}
\startdata
  A  &  03:44:12.80  & 32:01:37.10 & 091013 &  0.48 \\
  B  &  03:44:44.57  & 32:10:49.80 & 091013 &  0.51 \\
  C  &  03:44:23.45  & 32:01:49.39 & 091013 &  0.46 \\
  D  &  03:44:38.19  & 32:08:15.59 & 091020 &  0.80 \\
  E  &  03:44:35.91  & 32:09:04.09 & 091020 &  0.75 \\
  F  &  03:44:37.26  & 32:03:21.89 & 091020 &  0.77 \\
  G  &  03:44:30.31  & 32:09:47.49 & 091204, 091206 &  0.73 \\
  H  &  03:44:29.52  & 32:01:11.91 & 091204, 091206 &  0.68 \\
  I  &  03:44:40.71  & 32:09:56.90 & 091204, 091206 &  0.69 \\
  J  &  03:44:32.29  & 32:05:41.09 & 091207 &  0.62 \\
  K  &  03:44:13.64  & 32:13:27.60 & 091207 &  0.59 \\
  L  &  03:44:42.96  & 32:08:44.29 & 091207 &  0.61 \\
  M  &  03:44:37.60  & 32:11:57.59 & 091222 &  0.50 \\
  N  &  03:44:21.55  & 32:12:07.69 & 091222 &  0.49 \\
  O  &  03:44:26.39  & 32:08:12.49 & 091222 &  0.50 \\
  P  &  03:44:18.78  & 32:07:32.50 & 091227 &  0.46 \\
  Q  &  03:44:34.38  & 32:12:58.39 & 091227 &  0.48 \\
  R  &  03:44:45.34  & 32:04:01.79 & 091227 &  0.46 \\
  S  &  03:44:32.04  & 32:11:43.70 & 101006 &  0.94 \\
  T  &  03:44:43.53  & 32:07:42.74 & 101019 &  0.82 \\
  U  &  03:44:36.96  & 32:06:45.20 & 101122 &  0.91 \\
  V  &  03:44:56.15  & 32:09:15.19 & 101214 &  0.37 \\
\enddata
\end{deluxetable}

\begin{deluxetable}{rrrccccccc}
\tablecaption{Detected Disk Properties\label{tab:det}}
\tablenum{2}
\tablewidth{0pt}
\tablehead{\colhead{ID} & \colhead{R.A.} & \colhead{Decl} & \colhead{Field} & \colhead{Spectral Type} & \colhead{$F_{1.3}$} & \colhead{$\sigma_{1.3}$} & \colhead{$M_{d}$} & \colhead{$W_{\lambda}(H\alpha)$} & \colhead{$M_{star}$} \\ 
\colhead{} & \colhead{(J2000)} & \colhead{(J2000)} & \colhead{} & \colhead{} & \colhead{(mJy)} & \colhead{(mJy)} & \colhead{($M_{Jup}$)} & \colhead{(\AA)} & \colhead{($M_{\sun}$)} \\
\colhead{(1)} & \colhead{(1)} & \colhead{(1)} & \colhead{(2)} & \colhead{(3)} & \colhead{(4)} & \colhead{(5)} & \colhead{(6)} & \colhead{(7)} & \colhead{(8)}} 
\startdata
51    & 03~44~12.97 & 32~01~35.40 & A & \ldots & 75.90 & 0.55 & 69.19 & 45 & \ldots \\
153   & 03~44~42.77 & 32~08~33.90 & L & M4.75 & 6.49 & 0.71 & 5.92 & 40 & 0.20 \\
32    & 03~44~37.89 & 32~08~04.20 & D & K7 & 5.47 & 0.81 & 4.99 & 68 & 0.77 \\
221   & 03~44~40.26 & 32~09~33.20 & I & M4.5 & 3.23 & 0.89 & 2.94 & 40 & 0.20 \\
248   & 03~44~35.95 & 32~09~24.30 & E & M5.25 & 2.81 & 0.81 & 2.56 & 30 & 0.15 \\
468   & 03~44~11.07 & 32~01~43.70 & A & M8.25 & 2.37 & 0.56 & 2.16 & 400 & 0.02 \\
8078  & 03~44~26.69 & 32~08~20.30 & O & M0.5 & 2.07 & 0.51 & 1.89 & 75 & 0.65 \\
100   & 03~44~22.32 & 32~12~00.80 & N & M1 & 2.06 & 0.51 & 1.88 & 90 & 0.60 \\
192   & 03~44~23.64 & 32~01~52.70 & C & M4.5 & 2.06 & 0.57 & 1.88 & 40 & 0.23 \\
15    & 03~44~44.72 & 32~04~02.70 & R & M0.5 & 1.98 & 0.46 & 1.80 & 36 & 0.65 \\
\enddata
\tablecomments{(1) Source ID and coordinates as given in \cite{2006AJ....131.1574L}. (2) Observed field, as labeled in Figure \ref{fig:fields}. (3) Spectral type of object where available, from \cite{2003ApJ...593.1093L}. (4) $1.3$ mm flux density. (5) 1$\sigma$ errors in flux density.  (6) Protoplanetary disk mass in units of $M_{Jup}=1/1047M_{\sun}$. (7) H$\alpha$ equivalent widths, from \cite{2003ApJ...593.1093L}. (8) Stellar Masses estimated from pre-main-sequence evolutionary tracks.}
\end{deluxetable}

\clearpage

\begin{thebibliography}{42}
\expandafter\ifx\csname natexlab\endcsname\relax\def\natexlab#1{#1}\fi

\bibitem[{{Amelin} {et~al.}(2002){Amelin}, {Krot}, {Hutcheon}, \&
  {Ulyanov}}]{2002Sci...297.1678A}
{Amelin}, Y., {Krot}, A.~N., {Hutcheon}, I.~D., \& {Ulyanov}, A.~A. 2002,
  Science, 297, 1678

\bibitem[{{Andre} \& {Montmerle}(1994)}]{1994ApJ...420..837A}
{Andre}, P., \& {Montmerle}, T. 1994, \apj, 420, 837

\bibitem[{{Andrews} \& {Williams}(2005)}]{2005ApJ...631.1134A}
{Andrews}, S.~M., \& {Williams}, J.~P. 2005, \apj, 631, 1134

\bibitem[{{Andrews} \& {Williams}(2007)}]{2007ApJ...671.1800A}
---. 2007, \apj, 671, 1800

\bibitem[{{Andrews} {et~al.}(2011){Andrews}, {Wilner}, {Espaillat}, {Hughes},
  {Dullemond}, {McClure}, {Qi}, \& {Brown}}]{2011ApJ...732...42A}
{Andrews}, S.~M., {Wilner}, D.~J., {Espaillat}, C., {Hughes}, A.~M.,
  {Dullemond}, C.~P., {McClure}, M.~K., {Qi}, C., \& {Brown}, J.~M. 2011, \apj,
  732, 42

\bibitem[{{Arce} \& {Sargent}(2006)}]{2006ApJ...646.1070A}
{Arce}, H.~G., \& {Sargent}, A.~I. 2006, \apj, 646, 1070

\bibitem[{{Beckwith} {et~al.}(1990){Beckwith}, {Sargent}, {Chini}, \&
  {Guesten}}]{1990AJ.....99..924B}
{Beckwith}, S.~V.~W., {Sargent}, A.~I., {Chini}, R.~S., \& {Guesten}, R. 1990,
  \aj, 99, 924

\bibitem[{{Boss}(1997)}]{1997Sci...276.1836B}
{Boss}, A.~P. 1997, Science, 276, 1836

\bibitem[{{Carpenter}(2002)}]{2002AJ....124.1593C}
{Carpenter}, J.~M. 2002, \aj, 124, 1593

\bibitem[{{Carpenter} {et~al.}(2005){Carpenter}, {Wolf}, {Schreyer},
  {Launhardt}, \& {Henning}}]{2005AJ....129.1049C}
{Carpenter}, J.~M., {Wolf}, S., {Schreyer}, K., {Launhardt}, R., \& {Henning},
  T. 2005, \aj, 129, 1049

\bibitem[{{Cieza} {et~al.}(2007){Cieza}, {Padgett}, {Stapelfeldt}, {Augereau},
  {Harvey}, {Evans}, {Mer{\'{\i}}n}, {Koerner}, {Sargent}, {van Dishoeck},
  {Allen}, {Blake}, {Brooke}, {Chapman}, {Huard}, {Lai}, {Mundy}, {Myers},
  {Spiesman}, \& {Wahhaj}}]{2007ApJ...667..308C}
{Cieza}, L., {et~al.} 2007, \apj, 667, 308

\bibitem[{{Cieza} {et~al.}(2008){Cieza}, {Swift}, {Mathews}, \&
  {Williams}}]{2008ApJ...686L.115C}
{Cieza}, L.~A., {Swift}, J.~J., {Mathews}, G.~S., \& {Williams}, J.~P. 2008,
  \apjl, 686, L115

\bibitem[{{Currie} \& {Kenyon}(2009)}]{2009AJ....138..703C}
{Currie}, T., \& {Kenyon}, S.~J. 2009, \aj, 138, 703

\bibitem[{{D'Alessio} {et~al.}(2001){D'Alessio}, {Calvet}, \&
  {Hartmann}}]{2001ApJ...553..321D}
{D'Alessio}, P., {Calvet}, N., \& {Hartmann}, L. 2001, \apj, 553, 321

\bibitem[{{D'Antona} \& {Mazzitelli}(1997)}]{1997MmSAI..68..807D}
{D'Antona}, F., \& {Mazzitelli}, I. 1997, \memsai, 68, 807

\bibitem[{{Draine}(2006)}]{2006ApJ...636.1114D}
{Draine}, B.~T. 2006, \apj, 636, 1114

\bibitem[{{Eisner} \& {Carpenter}(2006)}]{2006ApJ...641.1162E}
{Eisner}, J.~A., \& {Carpenter}, J.~M. 2006, \apj, 641, 1162

\bibitem[{{Fedele} {et~al.}(2010){Fedele}, {van den Ancker}, {Henning},
  {Jayawardhana}, \& {Oliveira}}]{2010A&A...510A..72F}
{Fedele}, D., {van den Ancker}, M.~E., {Henning}, T., {Jayawardhana}, R., \&
  {Oliveira}, J.~M. 2010, \aap, 510, A72+

\bibitem[{{Feigelson} \& {Nelson}(1985)}]{1985ApJ...293..192F}
{Feigelson}, E.~D., \& {Nelson}, P.~I. 1985, \apj, 293, 192

\bibitem[{{Gomez} {et~al.}(1992){Gomez}, {Jones}, {Hartmann}, {Kenyon},
  {Stauffer}, {Hewett}, \& {Reid}}]{1992AJ....104..762G}
{Gomez}, M., {Jones}, B.~F., {Hartmann}, L., {Kenyon}, S.~J., {Stauffer},
  J.~R., {Hewett}, R., \& {Reid}, I.~N. 1992, \aj, 104, 762

\bibitem[{{Greaves} \& {Rice}(2010)}]{2010MNRAS.407.1981G}
{Greaves}, J.~S., \& {Rice}, W.~K.~M. 2010, \mnras, 407, 1981

\bibitem[{{Hatchell} {et~al.}(2005){Hatchell}, {Richer}, {Fuller},
  {Qualtrough}, {Ladd}, \& {Chandler}}]{2005A&A...440..151H}
{Hatchell}, J., {Richer}, J.~S., {Fuller}, G.~A., {Qualtrough}, C.~J., {Ladd},
  E.~F., \& {Chandler}, C.~J. 2005, \aap, 440, 151

\bibitem[{{Herbig}(1998)}]{1998ApJ...497..736H}
{Herbig}, G.~H. 1998, \apj, 497, 736

\bibitem[{{Hern{\'a}ndez} {et~al.}(2007){Hern{\'a}ndez}, {Hartmann}, {Megeath},
  {Gutermuth}, {Muzerolle}, {Calvet}, {Vivas}, {Brice{\~n}o}, {Allen},
  {Stauffer}, {Young}, \& {Fazio}}]{2007ApJ...662.1067H}
{Hern{\'a}ndez}, J., {et~al.} 2007, \apj, 662, 1067

\bibitem[{{Hubickyj} {et~al.}(2005){Hubickyj}, {Bodenheimer}, \&
  {Lissauer}}]{2005Icar..179..415H}
{Hubickyj}, O., {Bodenheimer}, P., \& {Lissauer}, J.~J. 2005, Icarus, 179, 415

\bibitem[{{Johansen} \& {Lacerda}(2010)}]{2010MNRAS.404..475J}
{Johansen}, A., \& {Lacerda}, P. 2010, \mnras, 404, 475

\bibitem[{{Kenyon} \& {Hartmann}(1995)}]{1995ApJS..101..117K}
{Kenyon}, S.~J., \& {Hartmann}, L. 1995, \apjs, 101, 117

\bibitem[{{Lada}(1987)}]{1987IAUS..115....1L}
{Lada}, C.~J. 1987, in IAU Symposium, Vol. 115, Star Forming Regions, ed.
  {M.~Peimbert \& J.~Jugaku}, 1--17

\bibitem[{{Lada} {et~al.}(2006){Lada}, {Muench}, {Luhman}, {Allen}, {Hartmann},
  {Megeath}, {Myers}, {Fazio}, {Wood}, {Muzerolle}, {Rieke}, {Siegler}, \&
  {Young}}]{2006AJ....131.1574L}
{Lada}, C.~J., {et~al.} 2006, \aj, 131, 1574

\bibitem[{{Luhman} {et~al.}(1998){Luhman}, {Rieke}, {Lada}, \&
  {Lada}}]{1998ApJ...508..347L}
{Luhman}, K.~L., {Rieke}, G.~H., {Lada}, C.~J., \& {Lada}, E.~A. 1998, \apj,
  508, 347

\bibitem[{{Luhman} {et~al.}(2003){Luhman}, {Stauffer}, {Muench}, {Rieke},
  {Lada}, {Bouvier}, \& {Lada}}]{2003ApJ...593.1093L}
{Luhman}, K.~L., {Stauffer}, J.~R., {Muench}, A.~A., {Rieke}, G.~H., {Lada},
  E.~A., {Bouvier}, J., \& {Lada}, C.~J. 2003, \apj, 593, 1093

\bibitem[{{Mann} \& {Williams}(2009)}]{2009ApJ...694L..36M}
{Mann}, R.~K., \& {Williams}, J.~P. 2009, \apjl, 694, L36

\bibitem[{{Mann} \& {Williams}(2010)}]{2010ApJ...725..430M}
---. 2010, \apj, 725, 430

\bibitem[{{Ormel} \& {Klahr}(2010)}]{2010A&A...520A..43O}
{Ormel}, C.~W., \& {Klahr}, H.~H. 2010, \aap, 520, A43+

\bibitem[{{Pollack} {et~al.}(1996){Pollack}, {Hubickyj}, {Bodenheimer},
  {Lissauer}, {Podolak}, \& {Greenzweig}}]{1996Icar..124...62P}
{Pollack}, J.~B., {Hubickyj}, O., {Bodenheimer}, P., {Lissauer}, J.~J.,
  {Podolak}, M., \& {Greenzweig}, Y. 1996, Icarus, 124, 62

\bibitem[{{Preibisch} \& {Zinnecker}(2002)}]{2002AJ....123.1613P}
{Preibisch}, T., \& {Zinnecker}, H. 2002, \aj, 123, 1613

\bibitem[{{Robitaille} {et~al.}(2006){Robitaille}, {Whitney}, {Indebetouw},
  {Wood}, \& {Denzmore}}]{2006ApJS..167..256R}
{Robitaille}, T.~P., {Whitney}, B.~A., {Indebetouw}, R., {Wood}, K., \&
  {Denzmore}, P. 2006, \apjs, 167, 256

\bibitem[{{Walawender} {et~al.}(2006){Walawender}, {Bally}, {Kirk},
  {Johnstone}, {Reipurth}, \& {Aspin}}]{2006AJ....132..467W}
{Walawender}, J., {Bally}, J., {Kirk}, H., {Johnstone}, D., {Reipurth}, B., \&
  {Aspin}, C. 2006, \aj, 132, 467

\bibitem[{{White} \& {Basri}(2003)}]{2003ApJ...582.1109W}
{White}, R.~J., \& {Basri}, G. 2003, \apj, 582, 1109

\bibitem[{{Williams} {et~al.}(2005){Williams}, {Andrews}, \&
  {Wilner}}]{2005ApJ...634..495W}
{Williams}, J.~P., {Andrews}, S.~M., \& {Wilner}, D.~J. 2005, \apj, 634, 495

\bibitem[{{Williams} \& {Cieza}(2011)}]{2011arXiv1103.0556W}
{Williams}, J.~P., \& {Cieza}, L.~A. 2011, ArXiv e-prints

\bibitem[{{Wilner} {et~al.}(2000){Wilner}, {Ho}, {Kastner}, \&
  {Rodr{\'{\i}}guez}}]{2000ApJ...534L.101W}
{Wilner}, D.~J., {Ho}, P.~T.~P., {Kastner}, J.~H., \& {Rodr{\'{\i}}guez}, L.~F.
  2000, \apjl, 534, L101

\end{thebibliography}
\end{document}